\documentstyle[sprocl]{article}

\input{psfig.sty}
\begin{document}

\title{INITIAL CHEMICAL ENRICHMENT IN GALAXIES}

\author{ Limin Lu\footnote{Hubble Fellow}, Wallace, L. W. Sargent, 
Thomas A. Barlow}

\address{Caltech, 105-24, Pasadena, CA 91125}


\maketitle\abstracts{ We present evidence that damped Ly$\alpha$
galaxies detected in spectra of quasars may not have started
forming stars until
the redshift $z\sim 3$.  If damped Ly$\alpha$ absorbers are the
progenitors of disk galaxies, then the above result may indicate 
that star formation in galactic disks first began at $z\sim 3$. }
  
\section{Introduction}

  One definition  of the epoch of galaxy formation is when galaxies first began
to form stars. Different types of galaxies (e.g., ellipticals and spirals) 
or different parts of the same type of galaxies (e.g., bulge and 
disk of spirals) may have formed at different epochs. A important goal of 
observational cosmology is to identify these different formation
epochs.

  Conventional wisdom suggests that ellipticals and the spheroidal
component of  spirals
formed very early, followed by disks. The population of galaxies
at $z>3$ identified using the Lyman limit drop-out
technique may very well be the progenitors of the spheroidal component
of massive galaxies [1,2]. 
Here we discuss evidence for a sharp rise in the metallicity 
distribution of damped Ly$\alpha$ absorbers at $z\leq 3$, 
which may signify the onset of star formation in galactic disks.

\section{Results}

   Damped Ly$\alpha$ (DLA) absorption systems seen in spectra of background
quasars are widely accepted to be the progenitors of present-day galaxies
[3], although their exact nature (dwarfs, spheroids, or disks?) is still 
unclear.  A program is carried out using the Keck telescopes to study the
chemical compositions of the absorbing gas in DLA systems.
One of the goals is to (hopefully) identify the epoch of the first episode
of star formation in these galaxies,
hence constraining theories  of galaxy formation.

   Figure 1 shows the distribution of [Fe/H] in DLA systems as a function
of redshift. Detailed descriptions of the data and analyses are given in
refs [4,5]. The low
metallicities of DLAs testify the youth of these galaxies: they have yet to
make the bulk of their stars. Remarkably, all 6 of the 
highest redshift absorbers
have [Fe/H]$\leq -2$; while many absorbers have reached ten times higher
metallicity at just slightly lower redshifts. This indicates an epoch of rapid
star formation at $z\sim 3$. The effect is likely to be real:
if DLA systems at $z>3$ have [Fe/H] that is uniformly distributed between
$-1$ and $-2.5$ (i.e.,, similar to the distribution at $2<z<3$), 
then the {\it posterior} probability for all six of the highest 
redshift systems to have [Fe/H]$\leq -2$ by chance is $1.4\times 10^{-3}$.

\begin{figure}
\psfig{figure=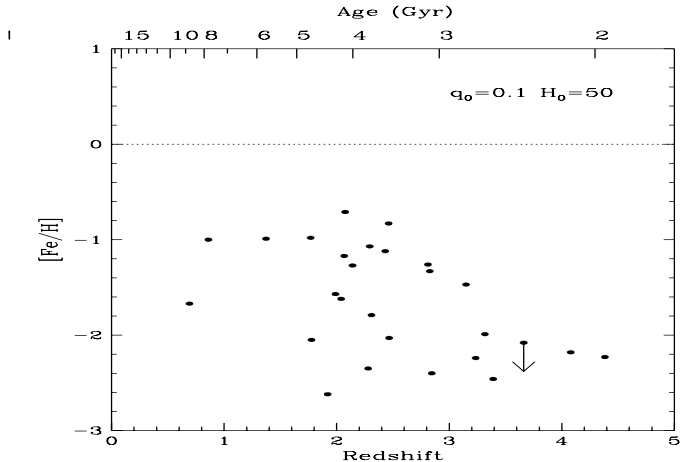,height=2.5in,width=3.5in}
\caption{Metallicity distribution of damped Ly$\alpha$ absorbers. 
   \label{fig:figure 1}}
\end{figure}

   Coincidentally, the metallicities of DLA systems at $z>3$, [Fe/H]$=-2$ to
$-2.5$, is identical (within the uncertainties) to those found for the 
IGM clouds at similar redshifts, as inferred from the C IV absorption 
associated with Ly$\alpha$ forest clouds [6,7,8].  This coincidence 
suggests that the metals in DLA galaxies at $z>3$ may simply reflect
those in the IGM, however they were made (e.g., Pop III stars, ejected from 
protogalaxies); DLA galaxies did not start making
their own stars (hence metals) until $z\sim 3$.

\section{Discussion}

The implications of the above result for the general question of
galaxy formation and evolution depend on the nature of the DLA galaxies.

   It was suggested [9] that DLA systems may represent the progenitors
of disk galaxies. This is supported by the very recent finding [10] that
the kinematics of DLA absorbers as inferred from the metal absorption line
profiles appears to be dominated by rotation with
large circular velocities ($>200$ km s$^{-1}$).
However, the mean metallicities of DLAs at $z>1.6$ are significantly 
below that of the Milky Way disk at the corresponding epoch [4,11].
The problem with the metallicity
distribution may be lessened if DLAs represent a thick disk phase of galaxies
[3] or if low surface brightness disk galaxies (which have substantially
sub-solar metallicities) make up a significant 
fraction of DLA absorbers [12]. {\it If} the
disk hypothesis for DLA absorbers is correct, we may have identified the
epoch of initial star formation in disk galaxies.

   Alternatively, DLAs may represent dwarf galaxies or the spheroidal 
component of massive
galaxies; this conjecture stems from  the similarity 
between the metallicity distribution of DLAs and those in 
halo globular clusters and local gas-rich dwarf galaxies [4]. 
In this case, however, one has to explain the kinematics of DLAs 
[10] by other means.

\section*{Acknowledgments}
LL appreciates support from a Hubble Fellowship (HF1062-01-94A). WWS
was supported by NSF grant AST95-29073.

\end{document}